\documentclass{article}
\usepackage{newcent}
\usepackage{bifchaos}
\usepackage{multicol}
\usepackage{vmargin}
\usepackage[dvips]{graphicx}

\setpapersize{A4}
\setmarginsrb{1.3cm}{2cm}{2cm}{2.5cm}{0pt}{0mm}{0pt}{0mm}

\begin{document}

\title{\bf FUZZY CONTROL OF CHAOS}

\author{OSCAR CALVO,\thanks{Email calvo@athos.fisica.unlp.edu.ar} 
\\
{\it CICpBA, 
L.E.I.C.I., Departamento de Electrotecnia, Facultad de Ingenier\'{\i}a,} \\   
{\it Universidad Nacional de La Plata, 1900 La Plata, Argentina} \\
\\
JULYAN H. E. CARTWRIGHT,\thanks{Email julyan@hp1.uib.es, WWW
http://formentor.uib.es/$\sim$julyan} 
\\
{\it Departament de F{\'{\i}}sica \& Centre de C\`alcul i Informatitzaci\'o,} \\
{\it Universitat de les Illes Balears, 07071 Palma de Mallorca, Spain} \\
}

\date{Int. J. Bifurcation and Chaos {\bf 8}, 1743--1747, 1998}

\maketitle

\begin{abstract}
We introduce the idea of the fuzzy control of chaos:
we show how fuzzy logic can be applied to the control of chaos, and provide an
example of fuzzy control used to control chaos in Chua's circuit.
\end{abstract}

\begin{multicols}{2}

\section{Introduction}
Chaos control exploits the sensitivity to initial conditions and to 
perturbations that is inherent in chaos as a means to stabilize unstable 
periodic orbits within a chaotic attractor. The control can operate by altering system 
variables or system parameters, and either by discrete corrections or by 
continuous feedback. Many methods of chaos control have been derived and 
tested \cite{chendong,lindner,ogorzalek}. Why then consider fuzzy control of 
chaos? 

A fuzzy controller works by controlling a conventional control method.
We propose that fuzzy control can become useful together with one of 
these other methods --- as an extra layer of control --- in order to improve 
the effectiveness of the control in terms of the size of the region over 
which control is possible, the robustness to noise, and the ability to control 
long period orbits.

In this paper, we put forward the idea of fuzzy control of chaos, and 
we provide an example showing 
how a fuzzy controller applying occasional proportional feedback to one of
the system parameters can control chaos in Chua's circuit.

\begin{figure*}[t]
\begin{center}
\includegraphics[width=0.7\textwidth]{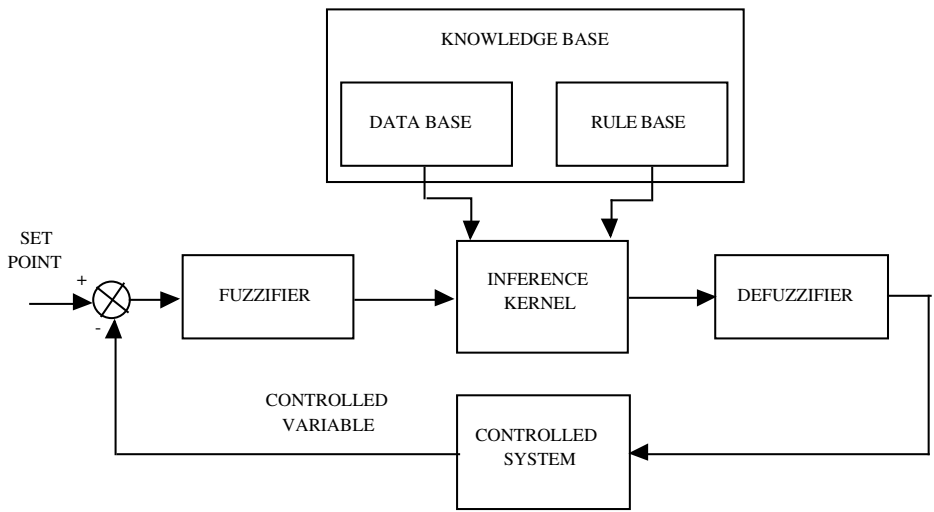}
\end{center}
\caption{\label{block_diag}
Fuzzy logic controller block diagram.}
\end{figure*}

\begin{figure*}[b]
\begin{center}
\includegraphics[width=0.6\textwidth]{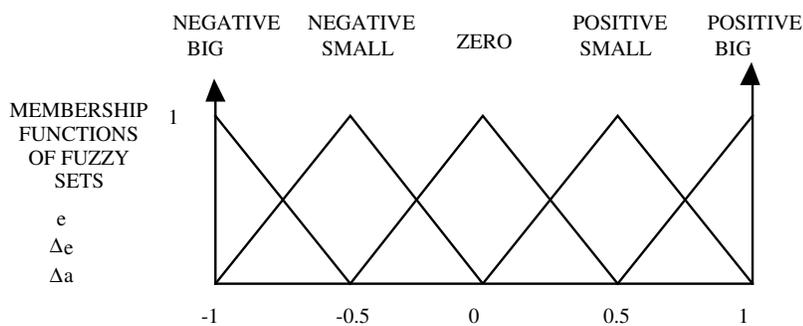}
\end{center}
\caption{\label{membership}
Membership functions of the input and output variables $e$, $\Delta e$, and 
$\Delta a$.}
\end{figure*}

\begin{figure*}[tb]
\begin{center}
\includegraphics[width=0.8\textwidth]{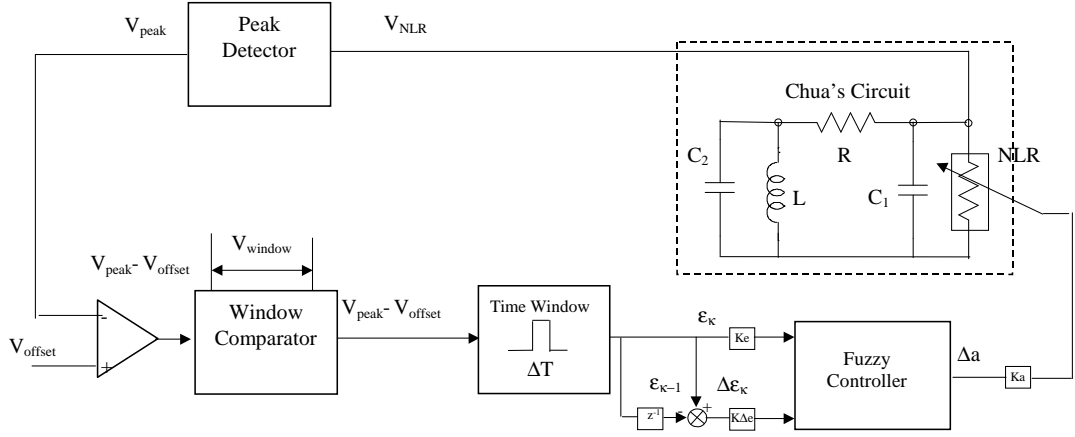}
\end{center}
\caption{\label{controller}
The whole controller and control system in the form of a block diagram, 
including the fuzzy controller, the peak detector, the 
window comparator, and the Chua's circuit system being controlled.}
\end{figure*}

\begin{table*}[tb]
\caption{\label{levels}
Quantification levels and membership functions.}
\begin{center}
\begin{tabular}{lccccccccc} 
\hline\hline
Error, $e$ & -1 & -0.75 & -0.5 & -0.25 & 0 & 0.25 & 0.5 & 0.75 & 1 \\ 
Change in error, $\Delta e$ & -1 & -0.75 & -0.5 & -0.25 & 0 & 0.25 & 0.5 & 0.75
& 1 \\
Control, $\Delta a$ & -1 & -0.75 & -0.5 & -0.25 & 0 & 0.25 & 0.5 & 0.75 & 1 \\ 
Quantification level & -4 & -3 & -2 & -1 & 0 & 1 & 2 & 3 & 4 \\
\hline\hline
Linguistic Variables & \multicolumn{9}{c}{Membership Functions} \\ \hline
Positive Big, PB & 0 & 0 & 0 & 0 & 0 & 0 & 0 & 0.5 & 1 \\
Positive Small, PS & 0 & 0 & 0 & 0 & 0 & 0.5 & 1 & 0.5 & 0 \\
Approximately Zero, AZ & 0 & 0 & 0 & 0.5 & 1 & 0.5 & 0 & 0 & 0 \\
Negative Small, NS & 0 & 0.5 & 1 & 0.5 & 0 & 0 & 0 & 0 & 0 \\
Negative Big, NB & 1 & 0.5 & 0 & 0 & 0 & 0 & 0 & 0 & 0 \\
\hline\hline
\end{tabular}
\end{center}
\end{table*}

\begin{table*}[tb]
\caption{\label{rules}
Rule table for the linguistic variables in Table~\ref{levels}.}
\begin{center}
\begin{tabular}{cc|ccccc} 
 & $e$ & NB & NS & AZ & PS & PB \\ 
$\Delta e$ & & & & & & \\
\hline
NB & & NB & NS & NS & AZ & AZ \\
NS & & NB & NS & AZ & AZ & PS \\
AZ & & NS & NS & AZ & PS & PS \\
PS & & NS & AZ & AZ & PS & PB \\
PB & & AZ & AZ & PS & PS & PB \\
\end{tabular}
\end{center}
\end{table*}

\begin{figure*}[tb]
\begin{center}
\includegraphics[width=0.75\textwidth]{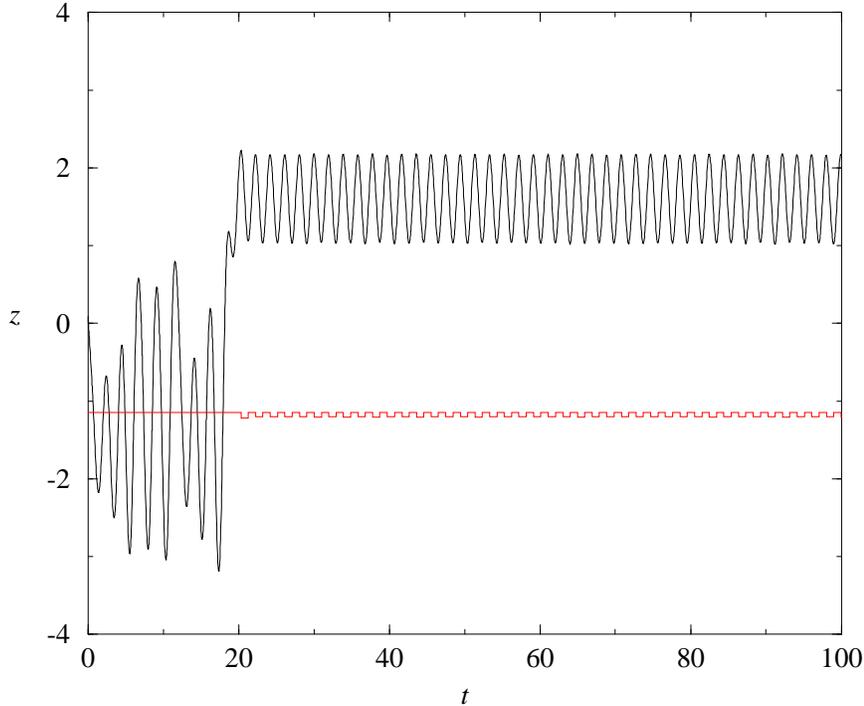}
\end{center}
\caption{\label{result}
The fuzzy controller stabilizes a previously unstable period-1 orbit.
The control is switched on at time 20. The lower trace shows the correction
pulses applied by the controller.}
\end{figure*}

\begin{figure*}
\begin{center}
\includegraphics[angle=-90,width=0.75\textwidth]{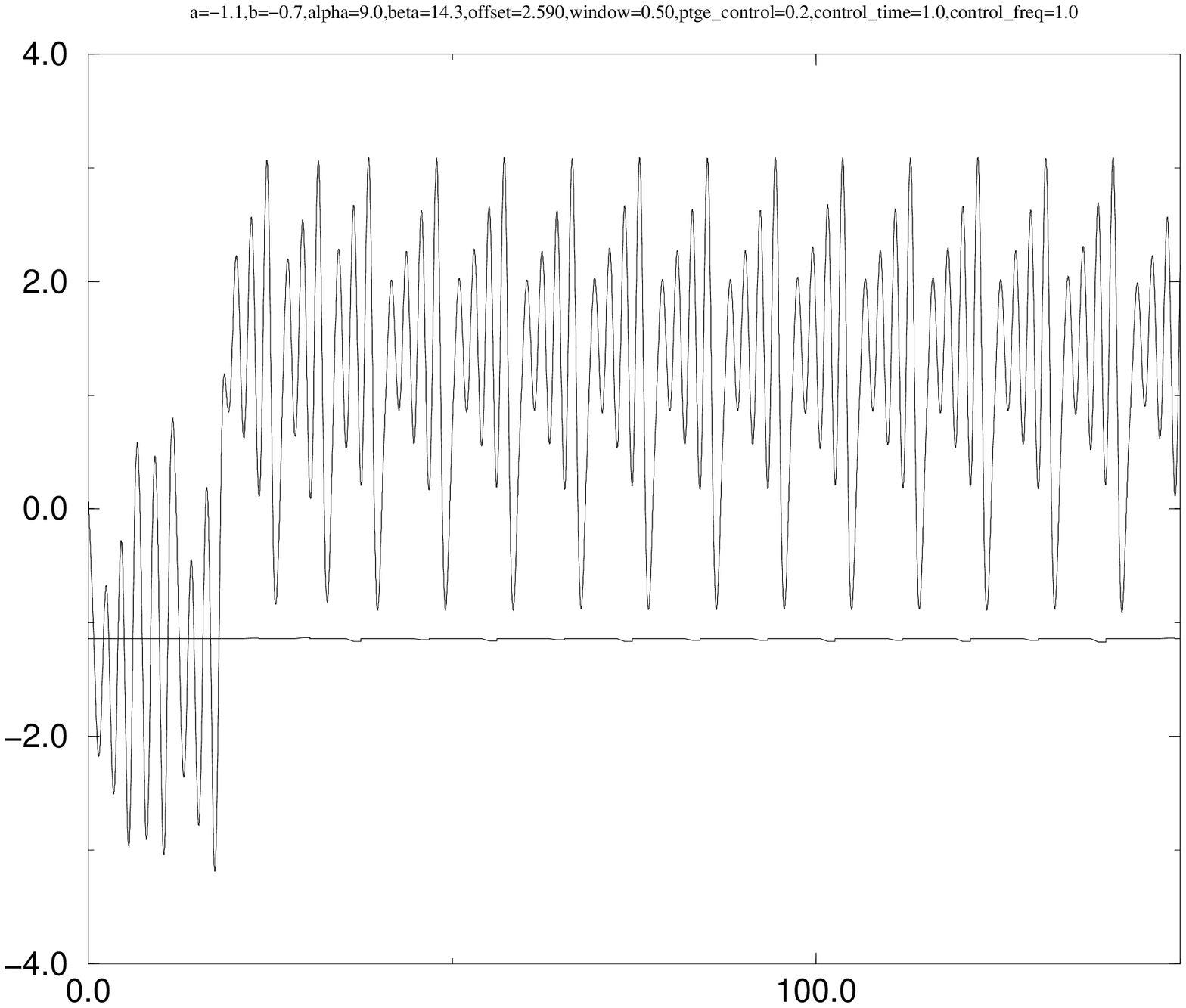} \\
\includegraphics[angle=-90,width=0.75\textwidth]{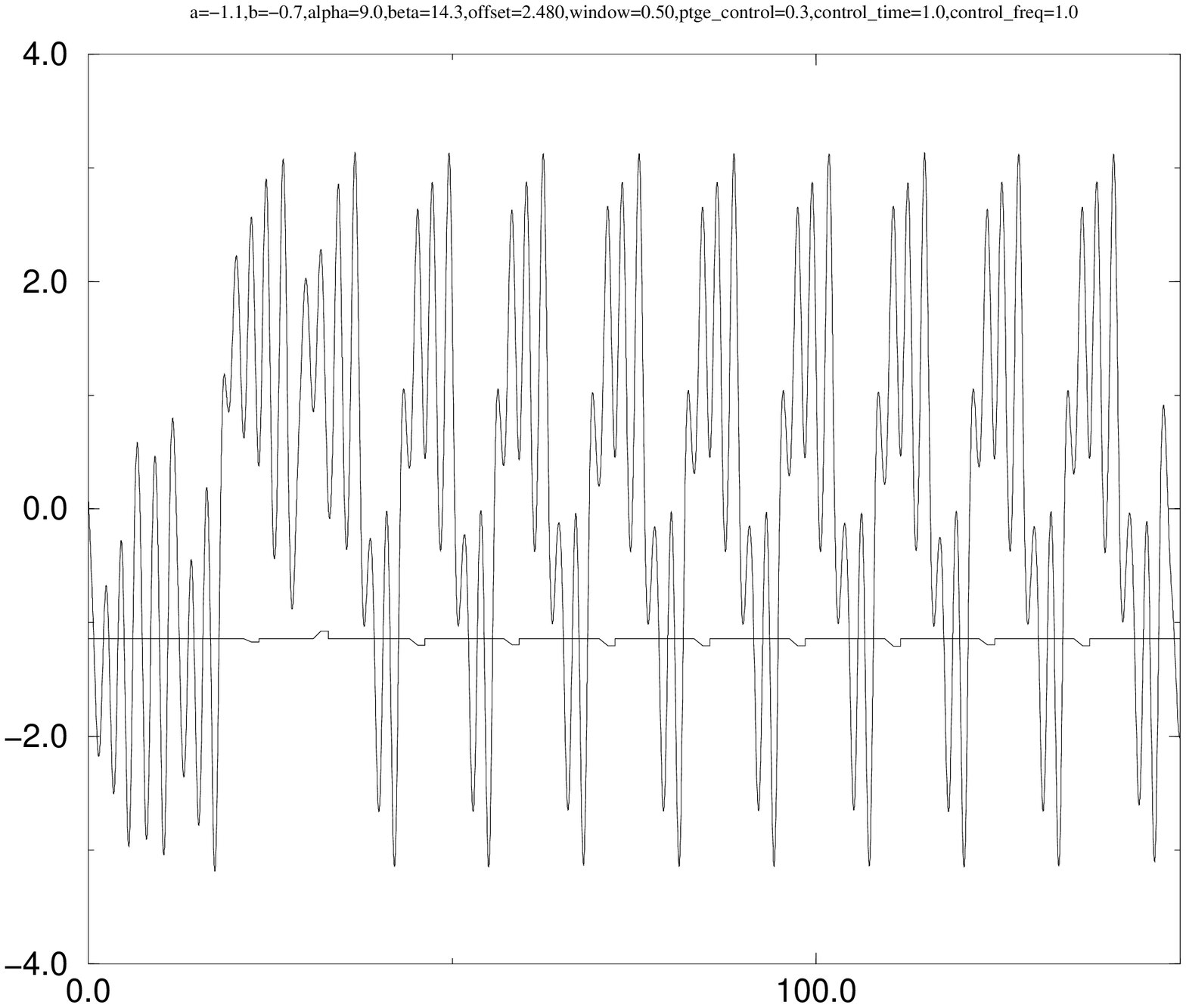}
\end{center}
\caption{\label{result2}
Trajectory traces show
higher period orbits stabilized by the controller. As before, the lower trace 
shows the correction pulses applied by the fuzzy control.}
\end{figure*}

\section{Fuzzy Control}
Fuzzy control \cite{driankov,terano} is based on the theory of fuzzy sets and 
fuzzy logic \cite{yager,bezdek}. The
principle behind the technique is that imprecise data can be classified 
into sets having fuzzy rather than sharp boundaries, which can be
manipulated to provide a framework for approximate reasoning in the face of
imprecise and uncertain information. Given a datum, 
$x$, a fuzzy set $A$ is said to contain $x$ with a degree of membership
$\mu_A(x)$, where $\mu_A(x)$ can take any value in the domain $[0,1]$. Fuzzy
sets are often given descriptive names (called linguistic variables) 
such as $\mbox{\em FAST}$; the membership function $\mu_{\mbox{\em FAST}}(x)$ 
is then used to reflect the similarly between values of $x$ and a contextual 
meaning of $\mbox{\em FAST}$. For example, if $x$ represents the speed of a car 
in kilometres per hour, and $\mbox{\em FAST}$ is to be used to classify cars 
travelling fast, then $\mbox{\em FAST}$ might have a membership function equal 
to zero for speeds below
90 km/h and equal to one for speeds above 130 km/h, with a curve joining
these two extremes for speeds between these values. The degree of truth
of the statement {\em the car is travelling fast} is then evaluated by
reading off the value of the membership function corresponding to the car's
speed.

Logical operations on fuzzy sets require an extension of the rules of classical
logic. The three fundamental Boolean logic operations, intersection, union,
and complement, have fuzzy counterparts defined by extension of the rules of
Boolean logic. A fuzzy expert system uses a set of membership functions and 
fuzzy logic rules to reason about data. The rules are of the form `if $x$ is 
$\mbox{\em FAST}$ and $y$ is $\mbox{\em SLOW}$ then $z$ is $\mbox{\em MEDIUM}$',
where $x$ and $y$ are input variables, $z$ is an output variable, and 
$\mbox{\em SLOW}$, $\mbox{\em MEDIUM}$, and $\mbox{\em FAST}$ are linguistic 
variables. The set of rules in a fuzzy expert system is known as the rule base, 
and together with the data base of input and output membership functions it
comprises the knowledge base of the system.

A fuzzy expert system functions in four steps. The first is
{\em fuzzification}, during which the membership functions defined on the
input variables are applied to their actual values, to determine the
degree of truth for each rule premise. Next under {\em inference}, the truth 
value for the premise of each rule is computed, and applied to the conclusion 
part of each rule.  This results in one fuzzy set to be assigned to each 
output variable for each rule. In {\em composition}, all of the fuzzy sets 
assigned to each output variable are combined together to form a single fuzzy 
set for each output variable. Finally comes {\em defuzzification}, 
which converts the fuzzy output set to a crisp (nonfuzzy) number.  

A fuzzy controller may then be designed using a fuzzy expert system 
to perform fuzzy logic operations on fuzzy sets representing 
linguistic variables in a qualitative set of control rules --- see 
Figure~\ref{block_diag}.

As a simple metaphor of fuzzy control in practice, consider the experience of 
balancing a stick vertically on the palm of ones hand. The equations of motion 
for the stick (a pendulum at its unstable fixed point) are well-known, but we 
do not integrate these equations in order to balance the stick. Rather, we 
stare at the top of the stick and carry out a type of fuzzy control to keep the 
stick in the air: we move our hand slowly when the stick leans by a small 
angle, and fast when it leans by a larger angle. Our ability to balance the
stick despite the imprecision of our knowledge of the system is at the heart 
of fuzzy control.

\section{Techniques for Fuzzy Chaos Control}
To control a system necessitates perturbing it.
Whether to perturb the system via variables or parameters depends on
which are more readily accessible to be changed, which in turn depends on what
type of system is to be controlled --- electronic, mechanical, optical, 
chemical, biological, etc. 
Whether to perturb continuously or discretely is a question of intrusiveness ---
it is less intrusive to the system, and less expensive to the controller, to 
perturb discretely. Only when discrete control is not effective might 
continuous control be considered. 

Ott, Grebogi, and Yorke \cite{OGY} invented a method of applying small feedback
perturbations to an accessible system parameter in order to control chaos. The
OGY method uses the dynamics of the linearized map around the orbit one wishes
to control. Using the OGY method, one can pick any unstable periodic orbit 
that exists
within the attractor and stabilize it. The control is imposed when the orbit
crosses a Poincar\'e section constructed close to the desired unstable periodic 
orbit. Since the perturbation applied is small, it is supposed that the 
unstable periodic orbit is unaffected by the control. 

Occasional proportional feedback \cite{hunt,lindner} is a variant of the 
original OGY chaos control method. Instead of using the unstable manifold of 
the attractor to compute corrections, it uses one of the dynamical variables,
in a type of one-dimensional OGY method. This feedback could be applied
continuously or discretely in time; in occasional proportional feedback it 
is applied discretely. Occasional proportional feedback exploits the strongly
dissipative nature of the flows often encountered, enabling one to control them 
with a one-dimensional map. The method is easy to implement, and in many cases 
one can stabilize high period unstable orbits by using multiple corrections per 
period. It is a suitable method on which to base a fuzzy logic technique for
the control of chaos, since it requires no knowledge of a system model, but 
merely an accessible system parameter. 

\section{An Example: Fuzzy Control of Chaos in Chua's Circuit}
Chua's circuit \cite{matsumoto,kennedy} 
exhibits chaotic behaviour that has been extensively studied, and whose 
dynamics is well known \cite{madan}. Recently, occasional proportional feedback 
has been used to control the circuit \cite{johnson}. The control used an 
electronic circuit to sample the peaks of the voltage across the negative
resistance and if it fell within a window, centred about $a$ by a set-point 
value, modified the slope of the negative resistance by an amount proportional 
to the difference between the set point and the peak value. The nonlinear nature
of this system and the heuristic approach used to find the best set of 
parameters to take the system to a given periodic orbit suggest that a fuzzy 
controller that can include knowledge rules to achieve periodic orbits may
provide significant gains over occasional proportional feedback alone.

We have implemented a fuzzy controller to control the nonlinearity of the
nonlinear element (a three segment nonlinear resistance) within Chua's circuit. 
The block diagram of the controller is shown in Figure~\ref{block_diag}. 
It consists of four blocks: knowledge base, fuzzification, inference and 
defuzzification. The knowledge base is composed of a data base and a rule base. 
The data base consists of the input and output membership functions
(Figure~\ref{membership}).  
It provides the basis for the fuzzification, defuzzification and inference
mechanisms.  The rule base is made up of a set of linguistic rules mapping 
inputs to control actions. Fuzzification converts the input signals $e$ and 
$\Delta e$ into fuzzified signals with membership values assigned 
to linguistic sets. The inference mechanisms operate on each rule, applying
fuzzy operations on the antecedents and by compositional inference methods
derives the consequents. Finally, defuzzification converts the fuzzy outputs 
to control signals, which in our case control the slope of the negative
resistance $\Delta a$ in Chua's circuit (Figure~\ref{controller}).  
The fuzzification maps the error $e$, and the change in the error $\Delta e$, 
to labels of fuzzy sets. Scaling and
quantification operations are applied to the inputs. Table~\ref{levels} shows 
the quantified levels and the linguistic labels used for inputs and output. 
The knowledge rules (Table~\ref{rules}) are represented as control statements 
such as `if $e$ is $\mbox{\em NEGATIVE BIG}$ and $\Delta e$ is 
$\mbox{\em NEGATIVE SMALL}$ then $\Delta a$ is $\mbox{\em NEGATIVE BIG}$'.

The normalized equations representing the circuit are 
\begin{eqnarray}
\dot x &=& \alpha \left(y-x-f(x)\right), \nonumber \\
\dot y &=& x-y+z, \nonumber \\
\dot z &=& -\beta y,
\end{eqnarray}
where $f(x)=b x+\frac{1}{2}(a-b) (|x+1|-|x-1|)$ represents the nonlinear 
element of the circuit. 
Changes in the negative resistance were made by changing $a$ by an amount
\begin{equation}
\Delta a =\mbox{Fuzzy Controller Output} \times \mbox{Gain} \times a.
\end{equation}

We have performed numerical simulations, both in C and in Simulink, of Chua's 
circuit controlled by the fuzzy logic controller. Figure~\ref{controller} shows
the whole control system in the form of a block diagram, including Chua's 
circuit, the fuzzy controller, the peak detector, and the window comparator. 
Figure~\ref{result} gives a 
sample output of the fuzzy controller stabilizing an unstable period-1 
orbit by applying a single correction pulse per cycle of oscillation.
By changing the control parameters we can stabilize orbits of different periods.
In Figure~\ref{result2} we illustrate more complex higher period orbits
stabilized by the controller.
One can tune the fuzzy control over the circuit to achieve the type of response
required in a given situation by modifying some or all of the rules in the
knowledge base of the system.

Of course, in the case of Chua's circuit the system equations are available and
fuzzy logic is thus not necessary for control, but this simple example permits 
us to see the possibilities that fuzzy control provides, by allowing a 
nonlinear gain implemented in the form of knowledge based rules.

\section{Conclusions}

We have introduced the idea of using fuzzy logic for the control of chaos.
Fuzzy logic controllers are commonly used to control systems whose dynamics 
is complex and unknown, but for expositional clarity here we have given an 
example of its use with a well-studied chaotic system. We have shown that it 
is possible to control chaos in Chua's circuit using fuzzy control. Further
work is necessary to quantify the effectiveness of fuzzy control of chaos
compared with alternative methods, to identify ways in which to systematically 
build the knowledge base for fuzzy control of a particular chaotic system, 
and to apply the fuzzy controller to further chaotic systems.

\bibliographystyle{bifchaos}
\bibliography{database}

\end{multicols}

\end{document}